\documentstyle[11pt,aaspp4]{article}
\def\ie{{\it i.e.,\ }}
\def\eg{{\it e.g.,\ }}

\def\etal{{\it et al.\ }}
\def\gtrapprox{\;\lower 0.5ex\hbox{$\buildrel >
    \over \sim\ $}}             
\def\lessapprox{\;\lower 0.5ex\hbox{$\buildrel < \over \sim\ $}}
\def\msol{\ifmmode {\>M_\odot}\else {$M_\odot$}\fi}
\def\pyr{\ifmmode {\>{\rm\ yr}^{-1}}\else {yr$^{-1}$}\fi}
\def\kms{\ifmmode {\>{\rm km\ s}^{-1}}\else {km s$^{-1}$}\fi}
\def\psqcm{\ifmmode {\>{\rm cm}^{-2}}\else {cm$^{-2}$}\fi}
\def\be{\begin{equation}}
\def\ee{\end{equation}}
\def\bea{\begin{eqnarray}}
\def\eea{\end{eqnarray}}
\def\tsigma{\ifmmode {\tilde\sigma}\else {$\tilde\sigma$}\fi}
\def\tomegao{\ifmmode {\tilde\omega_o}\else {$\tilde\omega_o$}\fi}
\def\tsmax{\ifmmode {\tilde\sigma_{i,{\rm max}}}\else
{$\tilde\sigma_{i,{\rm max}}$}\fi}
\received{Sept. 18, 1997}

\slugcomment{Submitted to The Astrophysical Journal Letters}

\lefthead{Maloney\& Begelman}
\righthead{Warped Disks in X-Ray Binaries}

\begin{document}

\title{The Origin of Warped, Precessing Accretion Disks in X-Ray Binaries}

\author{Philip R. Maloney\altaffilmark{1} and Mitchell
C. Begelman\altaffilmark{2,3}}

\altaffiltext{1}{Center for Astrophysics and Space Astronomy,
University of Colorado, Boulder, CO 80309-0389; maloney@shapley.colorado.edu}
\altaffiltext{2}{JILA, University of Colorado and National Institute
of Standards and Technology, Boulder, CO 80309-0440;
mitch@jila.colorado.edu} 
\altaffiltext{3}{Also at Department of Astrophysical and Planetary
Sciences, University of Colorado}

\begin{abstract}
The radiation-driven warping instability discovered by Pringle holds
considerable promise as the mechanism responsible for producing
warped, precessing accretion disks in X-ray binaries. This
instability is an inherently {\it global} mode of the disk, thereby
avoiding the difficulties with earlier models for the precession. Here
we follow up earlier work to study the linear behavior of the
instability in the specific context of a binary system. We treat the
influence of the companion as an orbit-averaged, quadrupole torque on
the disk. The presence of this external torque allows the existence of
solutions in which the direction of precession of the warp is
retrograde with respect to disk rotation, in addition to the prograde
solutions which exist in the absence of external torques. 
\end{abstract}


\section{Introduction}
For a quarter of a century, evidence has been accumulating for the
existence of warped, precessing disks in X-ray binary systems. The
discovery of a 35-day period in the X-ray flux from Her X-1
(\cite{tan72}) was interpreted almost immediately as the result of
periodic obscuration by a precessing accretion disk which is tilted
with respect to the binary plane  (\cite{kat73}). Katz
proposed that the precession was forced by the torque from the
companion star, but left unexplained the origin of the disk's
misalignment. An alternative possibility is the ``slaved disk'' model
of \cite{rob74}: here it is actually the companion star that
is misaligned and precessing; the accretion disk (fed by the
companion) will track the motion of the companion provided the
residence time in the disk is sufficiently short.

Dramatic evidence for a warped, precessing disk in an
X-ray binary was provided by the discovery of the relativistic
precessing jets in SS 433 (\cite{mar84}, and references
therein). The systematic velocity variations of the optical jet
emission, the radio jet morphology, and optical photometry of
the system all indicate a precession period for the disk of 164 days;
this must be a {\it global} mode, as both the inner disk (to explain
the jets) and the outer disk (to explain the photometry) must
precess at the same rate. The systematic variation of the X-ray
pulse profile of Her X-1 has been interpreted as the result of
precession of the inner edge of an accretion disk, which argues for a
global mode in this object also (\cite{tru86}; \cite{pet91}). A crucial
point is that in both Her X-1 and SS 433, the direction of precession
of the warp has been inferred to be retrograde with respect to the
direction of rotation of the disk (\eg \cite{ger76} for Her
X-1; \cite{leib84} and \cite{bri89} for SS 433).

A number of other X-ray binaries, both high and low mass, show
evidence for long period variations which may indicate the presence of
precessing, inclined disks: LMC X-4 (30.5 days), Cyg X-1 (294 days),
XB1820-303 (175 days), LMC X-3 (198 or 99 days), and Cyg X-2 (77 days)
(\cite{ph87}, and references therein; \cite{cow91}; \cite{sl92};
\cite{whi95}). Given the small number of X-ray binaries for which
adequate data exist to test for the existence of such periodicities,
it is evident that precessing, inclined accretion disks may be common
in X-ray binaries.

Although theoretical attempts to understand the mechanism responsible
for producing precessing, tilted or warped accretion disks date from
the discovery of Her X-1, no generally accepted model has emerged. The
original models suggested for Her X-1 suffer from serious flaws,
although both predict retrograde precession. The model of Katz (1973)
imposes a tilted disk as a boundary condition, but provides no
mechanism for producing this tilt. The slaved disk model (Roberts
1974) requires that the companion star rotation axis be misaligned
with respect to the binary plane; however, the axial tilt is expected
to decay by tidal damping on a timescale shorter than the
circularization time (\cite{chev76}). Other suggested mechanisms
suffer from the difficulty of communicating a single precession
frequency through a fluid, differentially-rotating disk (\eg
\cite{mb97}).

Recently, however, a natural mechanism for producing warped accretion
disks has been discovered. Motivated by work by \cite{pet77} and
\cite{ip90}, \cite{pri96} showed that centrally-illuminated accretion
disks are unstable to warping due to the pressure of re-radiated
radiation which, for a non-planar disk, is non-axisymmetric and
therefore exerts a torque. Further work on Pringle's instability was
done by \cite{mbp}, who obtained exact solutions to the linearized
twist equations, by \cite{mbn} (hereafter MBN), who generalized the
earlier work to consider non-isothermal disks (see below), and by
\cite{pri97}, who examined the nonlinear evolution. Radiation-driven
warping is an inherently global mechanism; the disk twists itself up
in such a way that the precession rate is the same at each radius.

Previous work on Pringle's instability assumed no external
torques. However, in X-ray binary systems the torque exerted on the
accretion disk by the companion star must dominate at large radii. In
the present paper we examine the behavior of the instability when an
external torque is included. In \S 2 we derive the modified twist
equation and solve it numerically, and in \S 3 we discuss the
solutions and the implications for X-ray binaries.

\section{The Twist Equation and Solutions}
We use Cartesian coordinates with the $Z-$axis normal to the plane of
the binary system; hence $Z=0$ is the orbital plane. As in MBN, we
assume that the disk viscosity $\nu_1=\nu_o\left(R/R_o\right)^\delta$,
where $R_o$ is an arbitrary fiducial radius. For a steady-state disk
far from the boundaries, this implies that the disk surface density
$\Sigma\propto R^{-\delta}$. In the $\alpha-$viscosity prescription,
$\delta=3/2$ corresponds to an isothermal disk. We distinguish between
the usual azimuthal shear viscosity $\nu_1$ and the vertical viscosity
$\nu_2$ which acts on out-of-plane motions (see \eg \cite{pp83},
\cite{pri92}); the ratio $\nu_2/\nu_1\equiv\eta$ is assumed to be
constant but not necessarily unity. Assuming an accretion-fueled radiation
source, we transform to the radius variable $x=(\sqrt
2\epsilon/\eta)\left(R/R_s\right)^{1/2}$, where $\epsilon\equiv L/\dot
M c^2$ is the radiative efficiency and $R_s$ is the Schwarzschild
radius. The linearized equation governing the disk tilt (including
radiation torque) for a normal mode with time-dependence $e^{i\sigma
t}$ is then
\be x{\partial^2 W\over \partial x^2}+
(2-ix){\partial W\over\partial x}=i\tsigma x^{3-2\delta} W.
\ee
(eq. [10] of MBN). The tilt is described by the function
$W\equiv \beta e^{i\gamma}$, where $\beta$ is the local tilt of the
disk axis with respect to the $Z-$axis and $\gamma$ defines the
azimuth of the line of nodes. We have 
nondimensionalized the eigenfrequency $\sigma$ by defining 
$\tsigma\equiv 2\eta^3 R_s^2\sigma/ \epsilon^4\nu_o$;
in general $\tsigma$ is complex, with real part $\tsigma_r$ and
imaginary part $\tsigma_i$. Negative values of $\tsigma_i$ correspond
to growing modes. We also define $R_o$ such that $x(R_o)=1$; therefore
$R_o=(1/2)(\eta/\epsilon)^2 R_s$.

Equation (1) assumes that the only torques are produced by viscous
forces and radiation pressure. In this case, the 
precession of the warping modes {\it must} be prograde, \ie in the
same direction as disk rotation (MBN). In a binary system, however,
the companion star also exerts a force on the accretion
disk. Since the orbital period is much shorter than the viscous
timescale, we average the resulting torque over azimuth and keep only
the leading (quadrupole) terms (\eg \cite{kat82}). This orbit-averaged
torque term will cause a ring of the disk that is tilted
with respect to the binary plane to precess, in a retrograde sense,
about the normal to the binary plane. (Including the time dependence
of the torque would result in nutation as well as precession, a
complication we do not consider here.) Relative to the angular
momentum density of a ring, $\Sigma R^2\Omega$, the torque term is 
\be 
{{\bf T}\over \Sigma R^2\Omega}={3\over
8}\left({GM\over r^3}\right)^{1/2} \left({R\over
r}\right)^{3/2}\left({M\over M_x}\right)^{1/2} \sin
2\beta\left(\sin\gamma,-\cos\gamma,0\right) 
\ee 
where $R$ is the distance from the compact object of mass $M_x$ and
$r$ is the separation between the companion star of mass $M$ and the
compact object. Writing the $R$-dependent coefficient of this
term as $\omega_o\left(R/R_o\right)^{3/2}$, the quadrupole torque then
contributes a term $i\tomegao x^3 W$ to the righthand side of the
twist equation (1), so that the equation becomes \be x{\partial^2
W\over \partial x^2}+ (2-ix){\partial W\over\partial
x}=ix^{3-2\delta}\left(\tsigma+\tomegao x^3\right)W\; , \ee where the
quadrupole precession frequency \tomegao\ has been nondimensionalized
in the same manner as the eigenfrequency in equation (1).

Equation (3) must be solved numerically. However, an entire class of
solutions can be derived from the results of MBN, which do
{\it not} include an external torque. Comparison of the twist equation
with and without quadrupole torque shows that equation (3)
with $\tsigma=0$ is formally identical to equation (1) (in which
$\tomegao=0$) if we replace $\delta$ with
$\delta'=\delta-3/2$. In other words, the purely precessing (\ie real
\tsigma, since \tomegao\ is real) modes with no external torque
have the same shapes as non-precessing modes {\it with}
quadrupole torque, with $\tomegao=\tsigma_r$, except that the latter
correspond to a larger value of the surface density index
$\delta$. The fact that these modes are non-precessing, \ie the warp
shape is fixed in an arbitrary inertial frame, is rather remarkable,
as it requires that the radiation torque (which attempts to make the
warp precess in a prograde sense) and the quadrupole torque (which
attempts to make the disk precess in the retrograde direction)
precisely balance at each radius.

We impose as the outer boundary condition that the disk must cross the
$Z=0$ plane at some radius, and impose the usual no-torque
inner boundary condition. This choice of $Z=0$ for the outer boundary
condition is not strictly correct; what we take as the disk boundary
here corresponds to the circularization radius, $R_{\rm circ}$, in a
real X-ray binary, and the actual outer boundary will typically be at
$R_{\rm out} \sim 3 R_{\rm circ}$. The proper boundary condition then
constrains the gradient $RW^{-1}\partial W/\partial R$ at $R_{\rm
circ}$ (J. Pringle 1997, private communication). We have modelled the
outer disk ($R_{\rm circ} \le R \le R_{\rm out}$), assuming that
$V_R=0$ for $R\ge R_{\rm circ}$ and $W'=0$ at $R_{\rm out}$, and have
examined how these solutions couple to the inner disk 
solutions. We find that the gradient is always steep at $R_{\rm
circ}$, so that the outer and inner disk solutions match up at radii
that differ by $\lessapprox 10\%$ from the radius of the zero, and the
tilt declines rapidly to zero for $R>R_{\rm
circ}$. Thus, the true solutions differ little from the
zero-crossing eigenfunctions calculated here, and we can ignore any
minor differences for the purposes of this paper. A full discussion of
this important point is given in MBN.

As in MBN, we separate equation (3) into real and imaginary parts and
solve as an initial-value problem, iterating to find the zero-crossing
eigenfunctions. In the absence of any external torque, there is a
marked change in the behavior of the eigenfunctions across $\delta=1$
(see MBN), so here we consider the two cases $\delta=0.75$
(corresponding to the usual gas pressure-supported Shakura-Sunyaev
disk) and $\delta=1.25$.

In Figure 1 we plot the location $x_o$ where the eigenfunction returns
to the $Z=0$ plane (in the terminology of MBN, these are the
first-order zeros, closest to the origin) and the normalized
precession rate $\tsigma_r$, as a function of the dimensionless
quadrupole frequency \tomegao, for $\delta=0.75$; Figure 2 plots the
same quantities for $\delta=1.25$. The series of curves are for
different growth rates, from a value of $\tsigma_i$ close to
the maximum growth rate, down to small growth rates for which the
curves essentially coincide with the steady-state $\tsigma_i=0$ curve,
not shown. The dotted portions of the curves mark the prograde modes,
the solid portions show the retrograde modes. Although there are
significant differences in behavior for the two values of $\delta$,
there are overall similarities: (1) There is a maximum value of
\tomegao\ above which the torque from the companion is too large to
allow warped modes to exist; (2) For a given value of \tomegao\ the
solutions are generally double-valued, with one prograde and one
retrograde mode or two prograde modes; the retrograde modes return to
the plane at larger radius than the prograde modes; (3) the precession
rates of the modes $\tsigma_r$ are much larger than the associated
values of \tomegao; this is unsurprising since the quadrupole torque
fixes the boundary condition at the disk's {\it outer} edge, which
is always at $R\gg R_o$ (recall $x(R_o)\equiv 1$).

For any single value of the growth rate $\tsigma_i$, the disk boundary
$x_o$ is at a fixed radius for a given value of \tomegao\ (and choice
of solution in the double-valued regime). The value of \tomegao, in
turn, is set by the masses and separation of the components of the
binary. In general this value of $x_o$ will not match the actual outer
boundary of the disk. This implies that in real disks, the location of
the outer boundary will determine the growth rate, provided that the outer
boundary falls within the range of radii occupied by the warp modes,
either prograde or retrograde. As can be seen from Figures 1a and 2a,
the fraction of the $(\tomegao,x_o)-$plane occupied by retrograde
modes is non-negligible, although it is smaller than that for
prograde modes. The retrograde modes essentially always have their
outer boundaries at larger radius than the prograde modes. The
precession rates $\tsigma_r$ are usually comparable for the prograde
and retrograde solutions, although for $\delta=1.25$, $|\tsigma_r|$ for
the retrograde modes may be several times larger than for the
prograde modes. (The abrupt termination of the curves for these
first-order eigenfunctions is genuine; the higher-order eigenfunctions
extend to larger $x$, generally with different values of $\tsigma_r$
and $\tsigma_i$ for a given value of \tomegao.)

\section{Discussion}
Where do real X-ray binary systems lie in this parameter space? In
order to quantify our results, we must make a specific assumption
about the viscosity; we also evaluate the results for $\delta=0.75$,
but there is in fact very little dependence on $\delta$. The
precession rates can be expressed in terms of the viscous timescale
$\tau_{\rm visc}\sim R/V_R\sim 2R^2/3\nu_1$ at the critical radius
$R_{\rm cr}$ (the minimum radius for instability: typically $R_{\rm
cr}\sim 10^9 \eta^2(0.1/\epsilon)^2(M/\msol)$ cm; see MBN), with
$x_{\rm cr}\equiv x(R_{\rm cr})$: 
\be
\sigma_r={\eta x_{\rm cr}^{4-2\delta}\over 12\tau_{\rm visc}(R_{\rm
cr})}\tsigma_r\,.
\ee
Using
the $\alpha-$prescription for viscosity, defining the precession
timescale to be $\tau_{\rm prec}\equiv 2\pi/\sigma_r$, and normalizing to
typical values, we find 
\be
\tau_{\rm prec}\sim 25 {\eta^2\over (\epsilon/0.1)^3}{M/\msol\over
(\alpha/0.1)} \left({H/R\over 0.01}\right)^{-2}_{R_{\rm cr}}\left(
{\tsigma_r\over 0.1}\right)^{-1}\;{\rm days\,.}
\ee
Thus the expected precession timescales for the disks are weeks to
months.

The quadrupole precession frequency $\omega_o$ is given by
\bea
\omega_o&=&{3\over 4}\left({GM\over r^3}\right)^{1/2} \left({R_o\over
r}\right)^{3/2}\left({M\over M_x}\right)^{1/2}\nonumber \\
&\simeq& 4.9\times 10^{-13}\left({\eta\over\epsilon}\right)^3
{\left(M/\msol\right)^2\over r_{11}^3}\left({M\over M_x}\right)^{1/2}
\;{\rm s}^{-1}
\eea
where $r_{11}=r/10^{11}$ cm. Since \tomegao\ is normalized in the same
manner as $\tsigma$, we find 
\be
\tomegao\sim 2\times 10^{-5} {\eta^5\over \epsilon_{0.1}^6}{\left(
M/\msol\right)^3\over(\alpha/0.1) r_{11}^3} \left({M\over
M_x}\right)^{1/2}\left({H/R\over 0.01}\right)^{-2}_{R_{\rm cr}}\;.
\ee
From Figure 1, it is evident that the relevant values of \tomegao\
fall in the range where warped disk solutions exist, with either
prograde or retrograde precession. Furthermore, the outer boundary
radius $x_o$ which characterizes these solutions is typically $R_{\rm
out}\sim 10^5 R_s$, which is also the expected value for real X-ray
binary sytems.

The estimates of $\tau_{\rm prec}$ and \tomegao\ are sensitive to the
value of the radiative efficiency $\epsilon$, extremely so in the
latter case. However, we do not expect this to be true in real X-ray
binary systems. The steep dependence on $\epsilon$ in the linear
theory estimate comes from the scaling of the fiducial radius
$R_o$. However, physically, the important quantity will be the value
of $\omega$ at the disk boundary, not its value at $R_o$, since it is
the outer boundary condition which determines the importance of the
torque from the companion. Thus we expect that in reality the
timescales and frequencies will be less sensitive to $\epsilon$ than
the above estimates.

The linear theory of disk warping leaves a number of questions
unanswered. Although we have demonstrated that retrograde as well as
prograde solutions exist when a quadrupole torque is included, there
does not appear to be any reason for choosing one mode over
another. For a fixed value of $\omega_o$, the retrograde solutions
(when both exist) occur for larger values of the boundary radius, but
this difference is not large, only a factor of $\sim 5-10$ at most
(see Figures 1 and 2). In Figure 3 (Plate 1) we show the shapes of the
most rapidly rotating prograde and retrograde modes for $\delta=1.25$,
for both the steady-state and fastest-growing modes. (The
$\delta=0.75$ modes are very similar; this difference from the
solutions with no external torque [MBN], in which the shapes of the
growing modes are very different for $\delta > 1$ and $\delta < 1$, is
a reflection of the importance of the quadrupole torque, which always
dominates at large radius.) The fast-growing prograde modes are more
``wound up'' (\ie the azimuth of the line of nodes rotates through a
larger angle between the origin and the disk boundary) than the
fast-growing retrograde modes, which may result in differences in the
effects of self-shadowing on the disk. An important observational
task, given our earlier argument that warped, precessing disks are
probably common in X-ray binaries, will be to determine if retrograde
precession is the norm in such systems, or only some fraction of
them. As we noted earlier, both of the original suggestions for
producing precessing disks in X-ray binaries require that the sense of
precession be retrograde.

We also cannot establish from linear theory that steady, long-lived
solutions actually exist. Work on the nonlinear evolution of
radiation-warped disks (with no external torques) by Pringle (1997)
suggests that chaotic behavior can result, in consequence of the
feedback between disk shadowing and the growth of the
warp. Calculation of the nonlinear evolution of disks in X-ray
binaries will be necessary to determine whether steady solutions do
exist in this case, and if so under what conditions. We have also
ignored the effects of winds, which could also drive warping
(\cite{sm94}; Pringle 1996). However, in the case of X-ray
heated winds, this seems unlikely to be important, since at the base of
the wind the gas pressure is much less than the radiation pressure
(\eg \cite{bms83}).

Pringle's instability appears to be very promising as a solution to
this long-standing problem. It is expected to be generically important
in accretion disks around compact objects, and is inherently a {\it
global} mode of the disk, so that the warp shape precesses with a
single pattern speed. In addition, it provides a natural mechanism for
producing non-planar disks in X-ray binary systems, so that the torque
from the companion star (which affects only the out-of-plane portion
of the disk) is able to influence the disk shape, allowing retrograde
as well as prograde modes to exist. 

\acknowledgements This research was supported by the Astrophysical
Theory Program through NASA grant NAG5-4061. MCB acknowledges support
from the NSF through grant AST-9529175. We are very grateful to Jim
Pringle for his insightful comments regarding the choice of outer
boundary condition, and to the referee for helpful and laudably prompt
remarks.

\clearpage

\figcaption[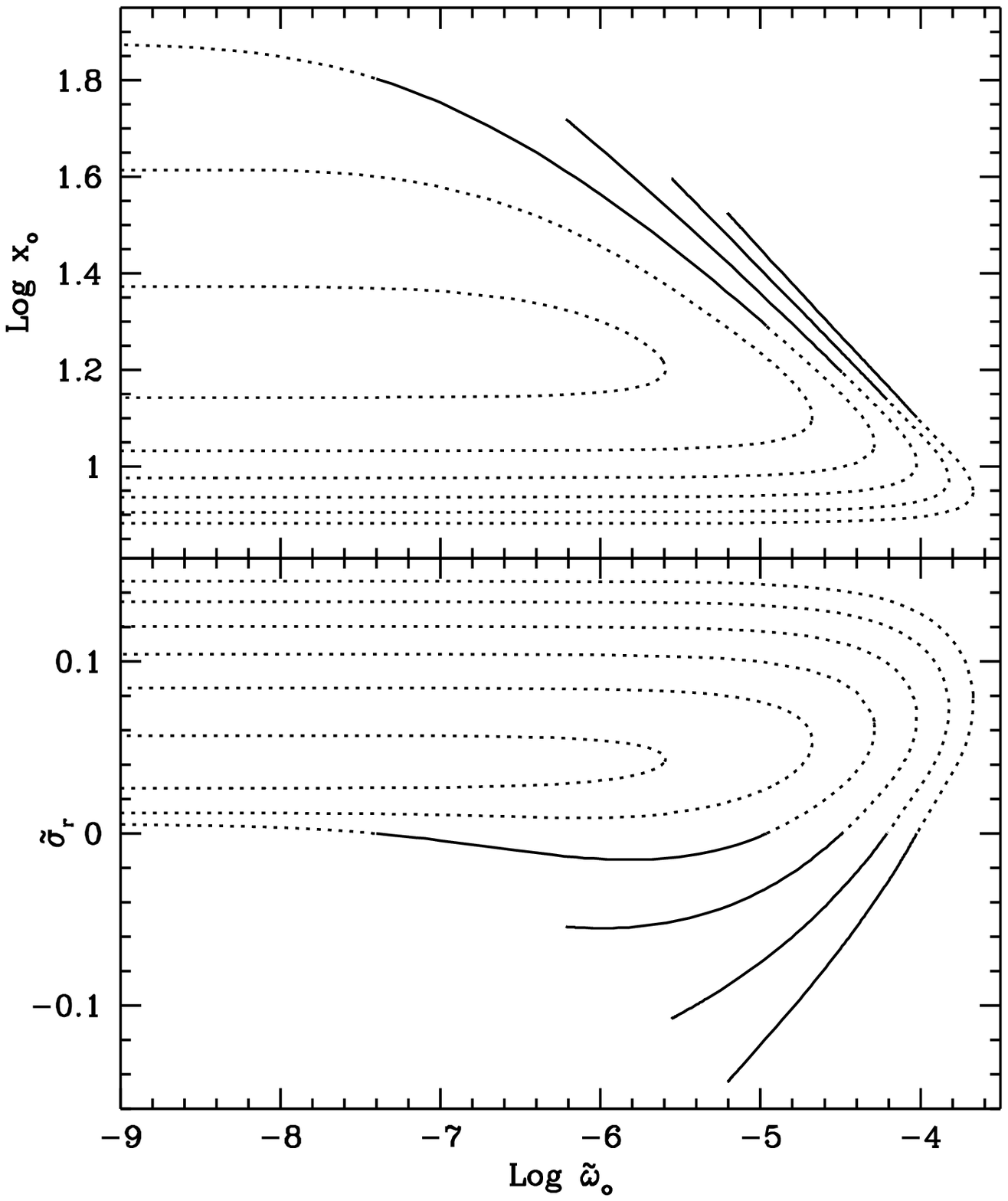]{(a) ({\it Top}) The location of the disk outer
boundary, $x_o$, as a function of the dimensionless quadrupole
precession frequency \tomegao, for surface-density index
$\delta=0.75$. From right to left, the curves are for increasing
values of the growth rate, $-\tsigma_i$; the plotted curves are for
$-\tsigma_i=0.001$, 0.01, 0.02, 0.03, 0.04, and 0.05. The solutions
with retrograde precession are marked with solid lines, the prograde
with dotted lines. The maximum growth rate plotted, 0.05, is very
close to the maximum allowed growth rate. (b) ({\it bottom}) The
dimensionless precession rate, $\tsigma_r$, for the same models as in
(a).}

\figcaption[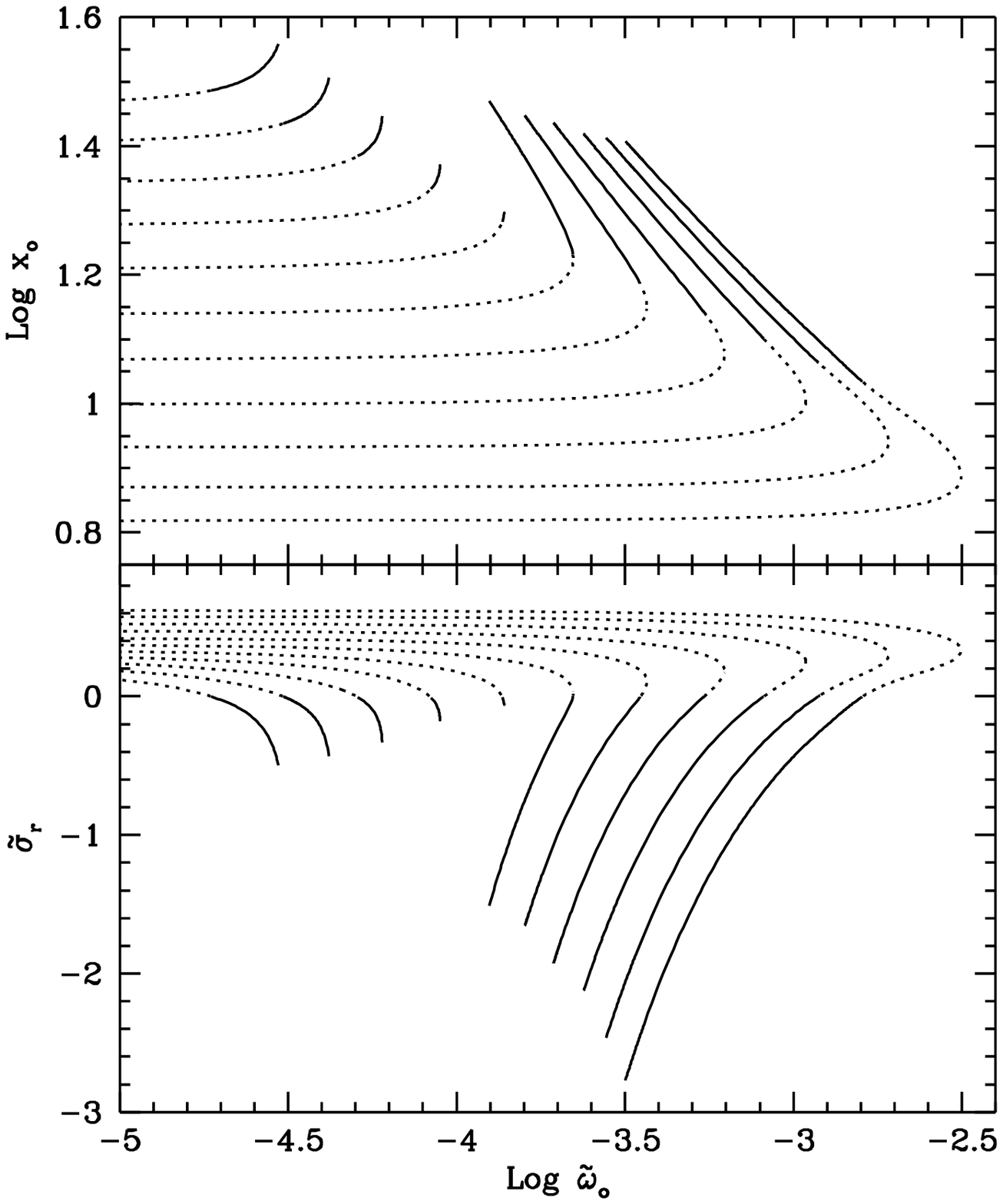]{(a) ({\it Top}) As in Figure 1a, for
$\delta=1.25$. From right to left, the plotted curves are for
$-\tsigma_i=0.01$, 0.1, 0.2, 0.3, 0.4, 0.5, 0.6, 0.7, 0.8, 0.9,
and 1.0. The maximum growth rate plotted, 1.0, is very close to the
maximum allowed growth rate. (b) ({\it bottom}) The dimensionless
precession rate, $\tsigma_r$, for the same models as in (a).}

\figcaption[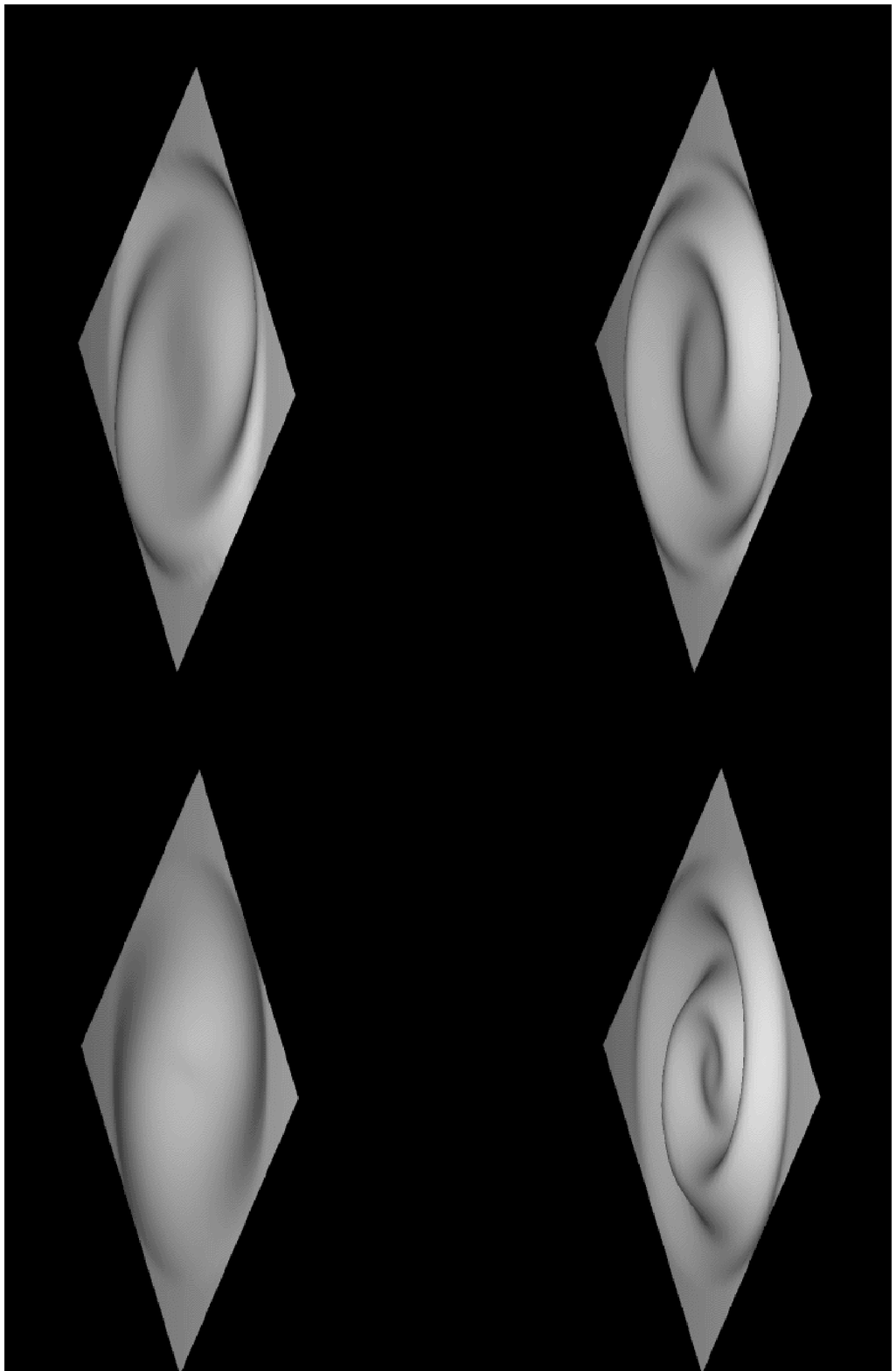]{Surface plots showing the shapes of several of
the warped disk modes, for $\delta=1.25$. In all cases the amplitude
of the warp has been fixed at $20\%$, and the solutions have been
plotted to the disk boundary. {\it Top left:} The most
rapidly-rotating (largest $\sigma_r$) steady-state ($\sigma_i=0$),
prograde mode. {\it Top right:} The most rapidly-rotating,
steady-state retrograde mode. {\it Bottom left:} The fastest-growing 
(maximum $-\sigma_i$), most rapidly-rotating prograde mode. {\it
Bottom right:} The fastest-growing, most rapidly-rotating retrograde
mode.}

\clearpage
\plotone{mb1.eps}

\clearpage
\plotone{mb2.eps}

\clearpage
\plotone{mb3.eps}

\end{document}